# Developers' Privacy Education: A game framework to stimulate secure coding behaviour *

*Note: Sub-titles are not captured in Xplore and should not be used

Abdulrahman Hassan Alhazmi
The Department of Computer Science
and IT at La Trobe University, Australia
and Jazan University, Saudi Arabia
a.alhazmi@latrobe.edu.au

Mumtaz Abdul Hameed
Technovation Consulting & Training
PVT
mumtazabdulhameed@gmail.com

Nalin Asanka Gamagedara Arachchilage
The School of Computer Science
The University of Auckland, New Zealand
and La Trobe University, Australia
nalin.arachchilage@auckland.ac.nz

*Abstract*—Software privacy provides the ability to limit data access to unauthorized parties. Privacy is achieved through different means, such as implementing GDPR into software applications. However, previous research revealed that the lack of poor coding behaviour leads to privacy breaches such as personal data breaching. Therefore, this research proposes a novel game framework as a training intervention enabling developers to implement privacy-preserving software systems. The proposed framework was empirically investigated through a survey study (with some exit questions), which revealed the elements that should be addressed in the game design framework. The developed game framework enhances developers' secure coding behaviour to improve the privacy of the software they develop.

## I. INTRODUCTION

Recently, there has been an increased use of software applications and the growth of the internet. As a result of that increase, lots of personal and sensitive information is generated among the users of the various programs [1]. This data is a major target to malicious software hackers, which leads to high data breaches in software systems [6], [10], [11]. One of the reasons for this might be that software developers are failing to implement laws in GDPR for applications. Researchers have indicated that even developers familiar with and have knowledge of the GDPR law still have issues implementing the law [2, 3]. GDPR has guidelines that enable software developers to implement privacy into software systems. It is also found that most developers do not understand the risk that exists if the GDPR law has not been adhered to. Software developers need tooling support or educational interventions due to the guidelines set by GDPR, which have been developed with a lawyer's mind rather than a software developer's. Therefore, it hinders developers from putting GDPR into practice and eventually leads to data breaches through the systems they develop [2].

GDPR outlines six important principles [4] that offer developers well-detailed and key guidelines. Privacy is protected by implementing these guidelines in the development of software systems [5]. The first principle is that data collection must be legitimate, translucent and impartial as defined in terms of Lawfulness, Transparency and Fairness [4]. The second principle is that there must be no hidden or wrong use of the client's data without his permission explained in the principle of Purpose Limitation [4]. The third principle, Data Minimization [4], states that only important and related data of the client that is required for the system must be collected. The fourth principle ensures safe storage and continuous modification of data, termed Accuracy [4]. Storage limitation is the fifth principle that ensures that unwanted or unrelated data must be deleted from the storage [4]. Integrity and Confidentiality is the last principle. Confidentiality states that only those with the requisite authorization are allowed to retrieve clients' data. Integrity, on the other hand, states that the client's retrieved data should only be altered by those authorized to carry out such alterations [4].

Although these guidelines exist, the occurrences of data breaches such as [6], [10], [11], and Yahoo [12] may indicate software developers do not adhere to the GDPR. Previous research has also shown that when developing any software system, the developers did not follow these principles to preserve the privacy of the system [6], [7], [8], [9] despite the existence of data privacy laws. Therefore, the main challenge to the developers can be said to be putting the data laws into practice. One of the top 5 issues in IT management is Data privacy [36]. It has also become one of the top 5 biggest issues in IT investments [10]. To ensure the security of information systems and protect organizations from security lapses, huge money is invested [9]. There is a continuous fight to eliminate these data breaches, but it is almost impossible, for example, the Zoom data breach [11]. Consistent data breaches and software application failure implies that using GDPR to protect the privacy of their clients can become a problem. The argument is made that this problem can be eliminated by educating and motivating software developers on the GDPR law, as they are neither privacy nor security experts [11]. This helps the developers to become a strong link which is able to avoid malicious mistakes, be accountable for their responsibility, understand their need to be aware of the risks involved in processing personal data and how to mitigate those risks and understand the work process. Takahashi and Kadobayashi [13] stated that the failure of a software developer is directly proportional to the failure of a cyber-security system that results in data breaches. According to Maasberg et al. 2016, the severity and frequency of malware attacks have increased [14]. Therefore, educating and motivating software developers on the implementation of privacy laws such as GDPR using gamification should be a priority [6], [9], [19]. A high level of Motivation will increase persistence, enhance the cognitive process and lead to improved performance. Software engineering education is important for developers to provide them with critical skills to meet the software industry's expectations and also help them to tackle problems. Game motivational aspects gain importance in Software Engineering Education because software engineers need to work in a highly social and collaborative environment, and games are good at providing the necessary collaboration while being played and designed [18].

Gamification refers to the art of adding game mechanics into nongame settings such as websites, online communities and learning management systems with the aim of enhancing participation [15]. Research by Subrahmanyam, Greenfield, and De Freitas, based on commercial games [16], suggests that digital games could encourage learning by creating a more entertaining and appealing environment. Games have also provided an inherently motivational approach to training and learning [16]. Gamification, in most case, revolve around creating a new environment around an existing activity by introducing elements that will increase the Motivation to do that activity. Gamification also ensures a level of security and privacy for the users' information while using gamified services [18]. Gamification increases learners' engagement, offers learners exciting, enriching learning environments, and allows designers' creativity to flourish. Gamification provides instant feedback whereby the learners can monitor their progress throughout the game and may even feel intrinsically motivated to complete the game successfully [43].

Security and privacy issues lead to threats of privacy on social media. The threats involved include data mining, phishing attempts, malware sharing, and Botnet attacks. Personal information serves as an attractive target for privacy breaches. Software developers need education to create operating systems, applications, and other computer software.

This research paper focuses on developing a gamification framework that will improve the developer's coding behaviour in terms of privacy embedment when developing software.

## II. BACKGROUND

Many researchers make an effort to evaluate the developers in the implementation of privacy-preserving software systems. Van et al. [23] have studied the attitudes of developers when they are managing the personal information of the clients according to the principles of GDPR. They also studied the developers when developing a privacy-preserving software system. They investigate how these developers implement the second principle of GDPR and use the data minimization technique. Researchers made a scale to measure the developer's attitudes when they manage their clients' personal information. Three main points are the primary objective of their study, i.e., informed consent, data minimization, and data monetization.

Studies investigated the factors that influence developers' behaviour and attitude in developing a privacy-preserving software system. Studies have focused on the privacy practices of organizations [20]. These studies give a deep understanding of the environmental factors that influence developers when managing privacy issues. It was recognized that organizations could implement an organization privacy environment regarding privacy issues to handle developers. Developers should have knowledge of secure development practices. Secure development practices help in reducing the vulnerability of privacy breaches and financial loss in companies and fatalities. The main Secure development practices include gamification, certifying developers, Save Developer Time, Accountability with Code, Offensive and Defensive approach, and Compliance [40]. This helps developers identify the most secure development practice to use regarding privacy issues.

Sheth et al. [21] studied the perception of users and developers towards privacy. A survey study was conducted to record the response towards privacy issues. The results show that users have become more careful about their personal data and location. They don't show much concern towards interaction data. The perception towards handling data differs depending on the region.

Concentrating on the contribution of software developers towards privacy implementation, Ayalon et al. [22] clarify that once there are established guidelines in place, developers do not follow the guidelines unless a systematic framework exists. Likewise, through an analytical investigation, Sheth et al. [21] discovered that developers find it difficult to recognize privacy guidelines and incorporate them into software systems. In the same way, Oetzel et al. [25] prove that developers need considerable effort from an end-user viewpoint to estimate privacy risks. In comparison, Ayalon et al. and Sheth et al. [21, 22] address the developers who face concerns and have trouble seeking to incorporate privacy into the information systems they are developing. Overall, these studies demonstrate the need for a systemic approach to direct software developers to integrate privacy into software systems that protect the privacy of users [24].

As evidenced in the literature review, many of the existing research and models did not consider developing educational interventions that will improve developers' coding behaviour. Without any empirical investigation, Arachchilage et al. [19] proposed a game-based teaching framework by combining a data minimization implementation model and a previously developed game design framework [18]. However, the framework did not incorporate all the GDPR elements – only the data minimization element. The proposed framework focuses on developing an education game-based framework through an empirical investigation that will improve developers' coding behaviour towards the implementation of privacy in software systems with the inclusion of all GDPR principles. With the ability of software engineering education, software developers will be able to identify problems, and with engineering skills, they will be able to solve the problem.

The framework Fig. 1 will incorporate all GDPR elements, which will be important through increasing the rights of data subjects by defining how organizations should process personal data [37]. It is through this definition by companies that users or data subjects are capable of trusting developers to process (i.e., analyze and implement) their data as per the guidelines of GDPR elements hence preserving the privacy of software systems. Lots of benefits can be pointed out in the incorporation of GDPR elements to the end-user and system's point of view. From the end-user perspective, GDPR elements will assist them through protection from unauthorized parties into their data, maintenance of the Accuracy of their personal data, being notified when their data is being used [38], protection of sensitive data and reservation of their personal data in the minimal duration. From the point of view of the system, GDPR elements have conditions in which it is necessary to inform the purpose of collecting personal information in easy-to-understand terms and to simplify the consent process [39], enabling transparency between the developer and the end-user, conservation of an authentic end-user data and it will automatically limit the users' data to the right parties. When the developer takes into account the GDPR elements in conjunction with the previously developed game design framework [18], they are able to protect the users' data and hence preserve privacy. Therefore, the developed game design framework assists the developers in improving their learning and applying the GDPR elements. In addition, the game design framework facilitates developers' secure coding behaviour through Motivation towards enabling privacy preservation in software applications.

Therefore, this research focuses on developing a game design framework through an empirical investigation to improve developers' coding behaviour using the application of mainly GDPR elements in order to mitigate the impact of data breaches in the software they develop.

## III. THE GAME DESIGN FRAMEWORK

Keep The game design framework, as shown in Fig. 1, begins with the presence of a threat and the developer who oversees preventing and controlling the threat from taking place. With the presence of a threat, the developer estimates (perceived susceptibility) the whole scenario of the threat towards the application and the level of the threat (perceived severity) will affect the application's user data if not handled well. After the identification of the estimates and threat level, the developers can now know (perceived threat) the kind of threat they are handling. The threat being known will force or stimulate (Motivation) the developer to come up with measures that will secure the users' data (secure coding behaviour). A successful application (safeguard effectiveness) will drive (Motivation) the developer to a higher point of controlling the breach or threat. The value (safeguard cost) of preventing the breach from occurring will inversely impact the developer since cost translates to limited resources. The urgency or compulsion (self-efficacy) towards the breach by the developers will be high for them to take charge of the threat. Additionally, GDPR principles can increase the developers' efforts towards preventing threats due to their great influence on securing the application. These elements will accelerate the developer's secure coding behaviour if well taken into consideration. When the threat is perceived after being estimated or suspected that it will happen, the developer will be motivated to take necessary actions according to the severity of the threat. The developer will ensure the application is safely safeguarded as they utilize fewer resources to have a favourable cost. The threat detected by the developer will drive the developer to have the capacity (Self-efficacy) to use their knowledge to deal with the breach.

The proposed framework, as shown in Fig. 1, contains seven hypotheses that are explained in the context of motivating software developers to enhance their coding behaviour towards privacy as follows: **H1**: Motivation affects developers positively towards developing privacy-preserving software. This means that if developers are stimulated more through several ways, they will be able to implement privacy in software effectively [28]. **H2**: Perceived threats affect developers positively towards developing privacy-preserving software. A well-detected threat will influence a developer to come up with strategies that block privacy breaches [28]. **H2a**: This is a combination of perceived threats and safeguard effectiveness that positively affects developers towards developing privacy-preserving software. If the developer can come up with an assumption that a threat will occur and realize that the effective solution will affect the software if it is not addressed, then they will be influenced to come up with measures to thwart the threat. **H3a**: Perceived severity affect developers positively towards developing privacy-preserving software. When a risk is realized by a developer that it will affect a given application if not addressed, then that will project the developer to strategize on the best means of controlling the risk or threat [28]. **H3b**: Perceived susceptibility affect developers positively towards developing privacy-preserving software. If a developer doubts that certain threats are going to affect the software, they will lay out positive steps to prevent the risk from taking place [28]. **H3c**: This is a combination of perceived susceptibility and perceived severity that positively affects developers towards developing privacy-preserving software. If the developer can come up with an assumption that a threat will occur and realize that the threat will affect the software if it is not addressed, then they will be influenced to come up with measures to thwart the threat [28]. **H4**: Safe-guard effectiveness affects developers positively towards developing privacy-preserving software. When a given application is robust, then the developer will be influenced to develop an application that adheres to privacy [28]. **H5**: Safe-guard cost affects developers negatively towards developing privacy-preserving software. This means that additional resources are needed for the preservation of privacy which is limited, leading to the hindrance of privacy preservation [28]. **H6**: Self-efficacy affects developers positively towards developing privacy-preserving software. A developer who can recognize a risk will have the desire to come up with means of preserving privacy in software and hence be influenced positively [28]. **H7**: The process of learning GDPR elements positively affects software developers' self-efficacy in developing privacy-preserving software systems [28]. When applying the six GDPR elements, it can be firmly observed that it will result in software developers enhancing their behaviour (through self-confidence) to integrate privacy into the software systems they develop.

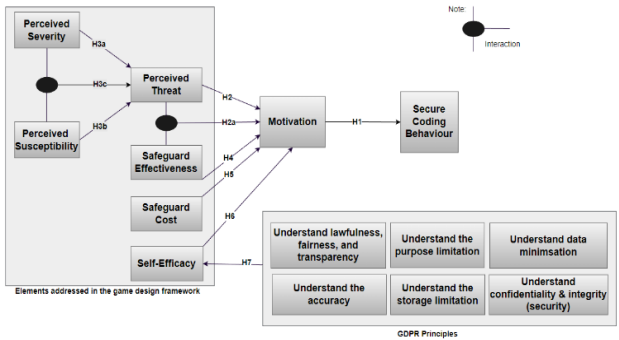

Fig. 1. The Proposed Serious Game Design Framework

## IV. METHODOLOGY

The main objective of this research is to elevate developers' skills through a game design framework through the implementation of GDPR principles to enable privacy-oriented software systems. The participants taking part were either professional developers or designers who were equipped with different levels of software skills. Individuals involved in the research were to give answers to questionnaires, and then an in-depth analysis of the answers was undertaken to assess the possibility of improving privacy in software applications. This research will follow both the quantitative and exit questions research methodology. Quantitative research methodology focuses on numbers or addresses the "what" or "how many" aspects of a research question [26]. Exit questions express the research data in terms of descriptive words focusing on characteristics. The quantitative approach was seen in the first closed-ended questionnaires. The data was collected directly from the developers, hence becoming primary data. Ground theory [29] was applied to analyze the answers to the last three exit questions.

### A. Ethics

The Ethics Board of our University gave approval for this project. Participants were asked to go through the consent form and provide their consent before starting the study. Those who failed to provide their consent to the study, we have not counted. We ensured that the data we collected from the participants were used and stored in a manner that guaranteed their privacy. Additionally, during our study, we did not request the participants to provide any sensitive information. We gave the participants an assurance that we would treat the information provided by them with the utmost Confidentiality. By and large, we abided by the GDPR while dealing with the data provided by the participants. Also, this investigation did not require personal publishing data. Finally, we told the participants that the results of the study would be brought to their knowledge.

### B. Pilot Study

A pilot study was carried out to determine whether the researchers should proceed with it and, if so, how [35]. A pilot study was done on a smaller scale than the main study, which means that the pilot study was necessary for the refinement of the quality and efficiency of the main study [35]. Ten participants (nine men and one woman) were recruited from different parts of the world by being sent an email message and being requested whether they could take part in the pilot study. Once they accepted, they were to answer the questionnaire provided. Developers should use secure practices such as the proposed game design framework to ensure security and privacy-preserving systems. Errors experienced at the time of the pilot study will result in refinements before the main study. The pilot study was conducted with 10 software developers. It involved questionnaire design, participants, procedures, results and summary.

*a) Questionnaire design*

The construction of the questionnaire was centred on previous game design frameworks [18] and GDPR principles [30]. The questionnaire items associated with this feature gauge developers' view of the likely damage that privacy breaches inflict. Perceived susceptibility [19] was employed to assess the chances and probability of the occurrence of a privacy breach. Perceived severity [19] was applied to display the extent to which a risk or threat, or harm can affect information privacy if the threat is not addressed. The items brought in perceived severity were centred around developers' concerns about the infringement of personal and confidential information on software applications [19]. The questions of safeguard effectiveness, safeguard cost, self-efficacy and GDPR were developed, majorly touching on secure coding behaviour and Motivation [18].

In that case, perceived threat holds three questions, perceived severity holds four items, perceived susceptibility holds three questions, safeguard effectiveness holds three items, safeguard cost holds three questions, self-efficacy holds three questions, Motivation holds three questions, secure coding behaviour holds three items and GDPR holds eight items. A five-point Likert scale at 1 = 'Strongly disagree' and 5 = 'Strongly agree' was applied and evaluated to the total 33 items.

*b) Participants*

The contributors to the pilot study questionnaire were ten individuals with different levels of software development skills from different regions of the world. An overview of the statistics of the participants in the pilot study is shown in Table 1.

TABLE 1. CONTRIBUTOR'S PILOT STUDY DEMOGRAPHIC

| Characteristics | Amount |
|---|---|
| Sample size | 10 |
| Gender | |
| Male | 9 |
| Female | 1 |
| Age (years) | |
| 18 - 24 | 3 |
| 25 - 34 | 6 |
| 35 - 44 | 1 |
| Average coding hrs (weekly) | |
| 11 – 15 | 1 |
| 21 – 25 | 2 |
| 26 – 30 | 3 |
| 31 – 35 | 1 |
| 36 – 40 | 3 |

*c) Procedure*

We have conducted 10 developers to complete quantitative and exit questions about their opinion on the educational model for privacy preservation of software. It was expected that it would take 20 minutes of their time. The questionnaire methodology will be used remotely to achieve diversity, recruit many participants, and curb the existing Covid-19 pandemic. After identification, questions and a consent form will be sent to the potential participants. Those agreeing to participate will check "I agree" in the participant information consent forms, which will be sent to them via Questionpro [27]. After a participant has agreed to participate in this research, they will complete the questions, which will include giving out their opinions on a 5-point Likert scale of the various questions. The answers to their questions measured the nine constructs: perceived severity, perceived susceptibility, perceived threat, safeguard effectiveness, safeguard cost, self-efficacy, Motivation, secure coding behaviour and GDPR. On the completion of the questionnaire and exit questions, participants were honoured for their precious time and effort in contributing to the study. The feedback is collected in a post-pilot survey [41]. A questionnaire is provided in the appendices.

*d) Results*

Cronbach's alpha, which is also referred to as a coefficient alpha, was used to measure the internal consistency of the questionnaire [31]. An alpha score that is greater than 0.7 indicates that there is a good ground of internal scale as dictated from previous research [31]. Ten individuals were required for the pilot study to take place, and therefore, Cronbach's alpha was calculated for each construct of the questionnaire and is summarized in Table 2.

TABLE 2. CRONBACH'S ALPHA SCORES FOR THE QUESTIONNAIRE CONSTRUCTS IN THE PILOT STUDY

| Constructs | Cronbach's alpha |
|---|---|
| Perceived susceptibility | 0.7954 |
| Perceived severity | 0.7615 |
| Perceived threat | 0.7591 |
| Safeguard effectiveness | 0.8189 |
| Safeguard cost | 0.8447 |
| Self-efficacy | 0.7752 |
| Motivation | 0.8072 |
| Secure coding behaviour | 0.7083 |
| GDPR | 0.7439 |

*e) Summary*

A slight revision was done on the degree of items of each construct, relying on the feedback extracted from the questions. Ten individuals were needed for the pilot study to take place. The final questionnaire contained three items for a perceived threat, four items for perceived severity, three items for perceived susceptibility, three items for safeguard effectiveness, three items for safeguard cost, three items for self-efficacy, three items for Motivation, three items for secure coding behaviour and eight items for GDPR. In that case, a total of 33 items were used in the pilot study to measure nine constructs in the research model using a five-point scale Likert at 1 = 'Strongly disagree' and 5 = 'Strongly agree'.

C. Main Study

Software developers have a direct impact on software development impact. Software engineering education provides students with the knowledge to transition to mature companies with defined structures in place successfully [42]. The participants were recruited from different parts of the world by being sent an email message and being requested whether they could take part in the main study. Once they accepted, they were to answer the questionnaire provided. Developers need to be educated to enhance secure coding behaviour, privacy-preserving software systems and secure software engineering to prevent a data breach.

*a) Participants*

In this study, a questionnaire was undertaken by 130 participants. These participants were developers and designers from different institutions. The age of the contributors varied from 18 to 65 or older, with 92 being male and 38 being female (as shown in Table 3). Their internet consumption and programming mostly ranged between 11-15 hours per week. The process was done remotely and discretionary. Table 3 below shows the demographics of the participants.

TABLE 3. CONTRIBUTOR'S MAIN STUDY DEMOGRAPHIC

| Characteristics | Amount |
|---|---|
| Sample size | 130 |
| Gender | |
| Male | 92 |
| Female | 38 |
| Age (years) | |
| 18 - 24 | 16 |
| 25 - 34 | 80 |
| 35 - 44 | 20 |
| 45 - 54 | 09 |
| 55 - 64 | 05 |
| Average coding hrs (weekly) | |
| 00 – 05 | 09 |
| 06 – 10 | 15 |
| 11 – 15 | 24 |
| 16 – 20 | 20 |
| 21 – 25 | 19 |
| 26 – 30 | 06 |
| 31 – 35 | 05 |
| 36 – 40 | 21 |
| 41+ | 11 |

*b) Procedure*

We have conducted 130 developers to complete quantitative and exit questions about their opinion on a GDPR educational model. A duration of around 20 minutes was needed for them to deal with the questions. The questionnaire methodology will be tackled remotely to achieve diversity, recruit many participants, and curb the existing Covid-19 pandemic. As soon as identification is finished, questions and a consent form will be forwarded to the potential participants. Individuals in agreement to take part will check "I agree" in the participant information consent forms, which will be dispatched to them via Questionpro [27]. Following the agreement of a participant to take part in this research, they will complete a number of questions which will include giving out their opinions on a 5-point Likert scale for closed-ended questions and the last open-ended questions. The answers to their questions measured the eight constructs: perceived severity, perceived susceptibility, perceived threat, safeguard effectiveness, safeguard cost, self-efficacy, Motivation and secure coding behaviour for the closed-ended questionnaires. At the end of the questionnaire, contributors were honoured for their precious time and effort in contributing to the study.

*c) Results*

In the previous analysis, that is, the pilot study, a calculation of Cronbach's alpha was carried out for every construct to evaluate the internal consistency of the questions in the research [31]. In Table 4, the outcome of the research of the 130 developers was outlined. A Cronbach's alpha above 0.7 in earlier studies indicates an internal consistency of a group of items [31].

Furthermore, the Kaiser–Meyer–Olkin (KMO) units are used to evaluate the fairness of the test. A number more than 0.6 for the KMO indicates an acceptable study to go ahead [32]. In this main study research, the KMO equals 0.853, hence acceptable.

TABLE 4. CRONBACH'S ALPHA SCORES FOR THE QUESTIONNAIRE CONSTRUCTS IN THE MAIN STUDY

| Constructs | Cronbach's alpha |
|---|---|
| Perceived susceptibility | 0.7816 |
| Perceived severity | 0.8605 |
| Perceived threat | 0.821 |
| Safeguard effectiveness | 0.9088 |
| Safeguard cost | 0.9239 |
| Self-efficacy | 0.8334 |
| Motivation | 0.7891 |
| Secure coding behaviour | 0.733 |
| GDPR | 0.7617 |

*d) Model testing*

Ling and Xue's theoretical model [33]. A theoretical model of technology threat avoidance theory (TTAT) was used to build the model [33], which describes how individual IT users avoid malicious information technology threats like spear-phishing. People's use of a particular safeguarding method is examined in the model [33]. There are many ways to protect yourself from phishing scams that don't necessitate the use of anti-phishing software.

In the study, a multiple regression analysis was applied to test Liang and Xue's theoretical model [33] using the following parameters: privacy breaches and education to curb privacy breaches and defence strategies in software applications.

The outcome for the model, as shown in Fig. 2, points out that 32.5%, 30%, and 29.2% of the variance are defined in perceived threat, Motivation, and secure coding behaviour, respectively. The strength and direction of the linear relationship between the two constructs were determined using Pearson correlation analysis. Indications from the analysis specify that perceived threat is notably dictated by perceived severity (r = 0.394**, and Sig. = .000) and perceived susceptibility (r = 0.319**, and Sig. = .000). Motivation is remarkably dictated by perceived threat (r = 0.38**, and Sig. = .000). Perceived severity and susceptibility on Motivation are highly brought about by perceived threat as stated by Liang and Xue's and Baron and Kenny's research [33]. It can be spotted that Motivation too, is ruled by self-efficacy (r = 0.369**, and Sig. = .000), safeguard cost (r = -0.036*, and Sig. = 0.683). and safeguard effectiveness (r = 0.386**, and Sig. = .000). Motivation is also crucially determined by GDPR (r = 0.434**, and Sig. = .000). In the end, secure coding behaviour (r = 0.54**, and Sig. = .000) is notably controlled by motivation. The interconnection between perceived severity and susceptibility was important to a perceived threat. The Motivation was remarkable through the interconnection between safeguard effectiveness and perceived threat.

Ground theory [29] was applied to analyze the answers to the last three open-ended questions. Answers that were based on a 'no or yes' were not handled by ground theory. Purposive sampling was done, and data collected through surveys relied on the three coding schemes, which were of much concern in detecting matters the developers come into contact with [34]. Open coding was the initial coding that we applied by reading the answers of both 10 and 130 developers severally and developing provisional markers to sum up the feedback [34]. Axial coding is the intermediate coding scheme used to point out the connection between open codes [34] for data saturation. Close answers in the axial coding stage were grouped together, and as a result, different groups were identified. Out of the different groups identified, the third coding scheme, advanced coding (selective coding), was applied to join the groups into minimal sets to examine perfectly [34] and for theoretical coding reaching theoretical saturation employed by theoretical sampling. The test results gave us a general idea that most developers shared the incapability of applying GDPR when embedding privacy into software applications. The findings based on ground theory from the question on the technique of improving secure coding behaviour when developing privacy-preserving software systems have seen that most of the developers would go for the option of choosing encryption as a means of preserving privacy in applications. The other common techniques involved training developers on the best security practices, legal expert engagement with developers to adhere to the GDPR policies mainly and constant interaction with the users of the software systems. Discoveries were seen in the answers of the developers in the query of the best training and education to increase secure coding behaviour, mainly touching on the use of learning materials such as videos, documentation, research papers and books. Extra means of training pointed out by the developers were the application of hacking games, carrying out workshops and engaging with skilled and professional developers. Most participants cited that they lacked enough Motivation to increase their secure coding skills. They also pointed out that time to improve secure coding behaviour was not enough. The constant evolution of software systems came out as a roadblock to achieving secure coding behaviour.

The test models, in general, held up all the hypotheses. Furthermore, education, gender, age and coding hours were among the control variables on secure coding behaviour and Motivation in the model testing, but the variables were noteworthy on the latter constructs.

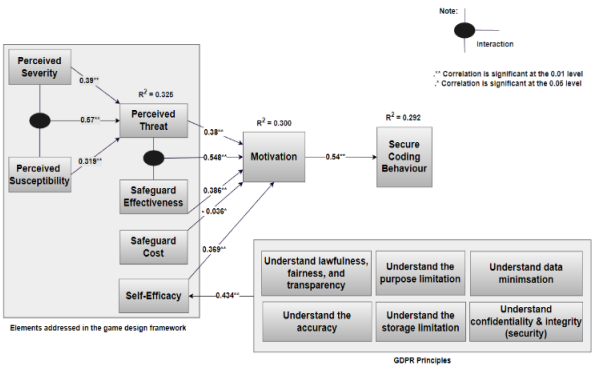

Fig. 2. Results of model testing

## V. SIGNIFICANCE OF GAME DESIGN FRAMEWORK

Avoidance of privacy breaches can be done by software developers by addressing the key elements in the previous game design framework and GDPR elements. The elements in the previous game design framework account for 32.5%, 30%, and 29.2% of the variance in perceived threat, Motivation, and secure coding behaviour, respectively. Perceived severity ($r = 0.394^{**}$, and Sig.= .000), perceived susceptibility ($r = 0.319^{**}$, and Sig. = .000) and their interaction ($r = 0.57^{**}$, and Sig. = .000) determine perceived threat. In that case, perceived severity and perceived susceptibility elements are marked in the previous game design framework for software developers to control privacy breaches. From Fig. 2, it can be seen perceived threat ($r = 0.38^{**}$, Sig. = 0.000), safeguard effectiveness ($r = 0.386^{**}$, Sig. = 0.000), safeguard cost ($r = -0.036^{*}$, Sig. = 0.683) and self-efficacy ($r = 0.369^{**}$, Sig. = .000) in connection with GDPR principles, determine secure coding behaviour of developers. Nevertheless, it is important to point out that safeguard cost affects Motivation negatively, but it is majorly ruled by Motivation. The reason for being ruled by Motivation is that the developer's desire to keep away from the breach is anticipated to be diminished by the cost of applying the safeguard measure [33]. Hence, safeguard effectiveness, perceived threat, safeguard cost, self-efficacy elements and the six GDPR principles should be taken note of in the new game design framework. In the end, secure coding behaviour ($r = 0.54^{**}$, and Sig. = .000) is influenced by Motivation.

On that account, perceived susceptibility, safeguard effectiveness, perceived threat, safeguard cost, perceived severity and self-efficacy elements, together with the six GDPR principles, are all addressed in the game design framework, as shown in Fig. 3.

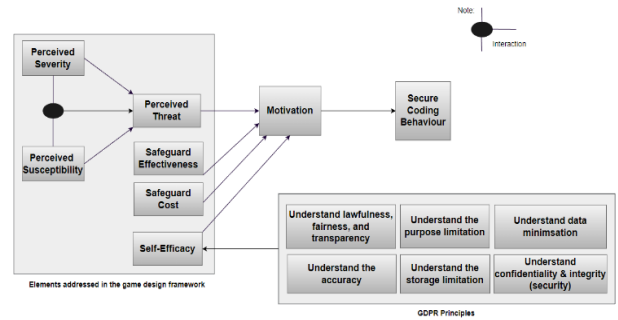

Fig. 3. Game design framework and GDPR principles

## VI. DISCUSSION

The aim of this research is to look through gamification for software developers to control and prevent privacy and data breaches. In that case, privacy and data breaches and anti-breaching learning of privacy and data were taken as a security vulnerability and safeguarding measure, respectively. Due to their high impact on software developers' secure coding behaviour, threat perception was highly considered in the research. The research results in Fig. 2 indicate a reasonable value of variance in Motivation and secure coding behaviour of 30% and 29.2%, respectively. In general, this research passes a clear and fine idea that we have to motivate developers to stay away from privacy and data breaches.

An important element to take note of is Motivation which has a low variance (30%). Due to it being low, there is a clear clarification for that outcome. Whenever a developer takes any step towards the data or privacy breach they encounter, they are driven in a number of different ways to prevent and take control of the situation. When the breach can be prevented, the developers have driven automatically through intrinsic motivation [17]. Nonetheless, when the breach cannot be prevented, the developers will be driven by extrinsic Motivation. If the breach can be controlled and prevented by an outside developer, the developer will be driven through motivation [17]. The three types of motivations [17] could increase a lot towards the prevention and control of privacy and data breaches. Intrinsic Motivation is defined as the drive to perform something in the absence of a clear outside award; for example, pleasure or happiness [17]. There is a clear relationship between intrinsic Motivation and privacy breaches through the following protective actions. Whenever developers face a privacy breach that is within their expertise, they are usually intrinsically motivated to deal with the breach. Extrinsic Motivation is explained as the efforts that are propelled by external rewards, rewards such as praise, fame, and money [17]. A breach that is not within the developer's skillset will drive him or her to be extrinsically motivated by rewarding them with things like money to drive them to study and added knowledge for safeguarding measures. Motivation drive is referred to as a non-self-determined effort towards a task which means an individual may or may not perform the act without any intentions. If the privacy breach prevention measures are beyond the developer's skillset and research, they will be driven through Motivation. The developer will seek external assistance towards the privacy breach. Generally, it can be declared in the current research that developers will experience different types of motivations towards the implementation of secure coding behaviour on privacy and data breaches.

Software developers need to be assured that privacy and data breaches are inevitable in the software environment and can be prevented and controlled. The analysis of the variance in perceived threat (32.5%) indicates that the value is reasonable. On that account, threat perception is important in the gamification framework. Additionally, the research indicates that the threat perception of developers has to be conscious of the probability and severity of privacy and data breach. When developers recognize the breach, they will be driven to stay away from it. Safeguard effectiveness, cost and confidence are the three main aspects of safeguard measures. A high motivation for the prevention of privacy breaches by developers will come up because of a high degree of safeguard effectiveness. High confidence in carrying out safeguarding measures impacts the Motivation of developers towards privacy and data breaches. In that case, self-efficacy has a great advantage in the framework to refrain from breaches through Motivation.

A higher cost of protecting applications translates to a lower motivation of developers towards privacy breach prevention. The research from [33] outlines that when capital, time, disruptions, and apprehension required to apply safeguarding measures are excessive, developers will be less motivated to prevent and control breaches [33]. The recent results indicate that safeguard cost negatively influence motivation [28]. On that account, safeguard cost should be taken note of for the application to be effective. The six GDPR principles are another key element in the implementation of secure coding behaviour among software developers through Motivation to avoid privacy breaches. When the GDPR principles are implemented in the gamification framework alongside the previous game design framework, it will result in software systems that meet secure coding behaviour among software developers through Motivation. The research disclosed that perceived threat is highly influenced by the interaction between perceived susceptibility and severity ($r = 0.57**$, sig. $= .000$)

This research stresses that Motivation is notably influenced by the communication of perceived threats and safeguarding effectiveness in conjunction with six GDPR principles. The communication of the two elements can be categorized into two standpoints. First, when the privacy breach degree is high, a perceived threat can be seen to negatively affect the association between motivation and safeguard effectiveness. Finally, if the degree of safeguard effectiveness is high, it is seen to negatively affect the association between perceived threat and Motivation.

In summary, this research articulates that developer-centric security training and educational interventions should be designed in a way that the player perceives a threat (i.e., privacy threats in our case), for example, through a game scenario/story. It then extrinsically motivates developers to enhance their confidence (i.e., self-efficacy) through learning how to write secure code to embed privacy into the software. Finally, the teaching content (i.e., GDPR elements in our case) can be developed and incorporated into a training and educational platform (i.e., a game) to learn and build his confidence, leading to nudge their behaviour (i.e., secure coding behaviour).

VII. CONCLUSION AND FUTURE WORK

Privacy-preserving software applications through a game design framework as a training mechanism to increase secure coding behaviour (i.e., to develop privacy-preserving software systems) is the core proposition of this research. Through an empirical investigation, the proposed game design framework nudges developers' secure coding behaviour to prevent privacy breaches in applications. Extraction of the past game design framework and the GDPR principles attain the aim of this research. The game player (developer) can gain confidence (i.e., self-efficacy) and information to produce applications that preserve the end-users privacy.

Further study should be conducted to experimentally investigate the suggested game design framework via software developers to examine their (secure) coding behaviour. Gaming prototypes may be conceived (through storyboarding) and produced (via both low and high-fidelity application prototypes) using a variety of design methodologies, such as participative or scenario-based game design.

An experimental technique, such as a think-aloud study, can be used to investigate the influence of participants on the developed game design framework (i.e., secure coding behaviour) following their interaction with the developed game prototype.

We want to undertake the following in the future:

- Developing a game prototype that combines elements from the framework we established - this will be accomplished through focus groups.
- Putting the game prototype into action on a browser-based platform.
- User testing - whereas users may be interested in the application, we will evaluate their effect on the theoretical framework developed.
- Measuring how our game improves software developers' secure coding behaviour - that is, in terms of designing privacy-preserving software systems.
- Conducting collective longitudinal studies to assess knowledge retention.